\date{}
\documentclass[showpacs]{revtex4}
\usepackage{graphicx,psfrag,amsmath,amssymb,amsfonts,bbm,latexsym,color,dcolumn,epsf}

\begin{document}
\title{Noise-induced energy excitation by a general environment}

\author{Fernando C. Lombardo \footnote{lombardo@df.uba.ar}}
\author{Paula I. Villar \footnote{paula@df.uba.ar}}
\affiliation{Departamento de F\'\i sica {\it Juan Jos\'e
Giambiagi}, FCEyN UBA, Facultad de Ciencias Exactas y Naturales,
Ciudad Universitaria, Pabell\' on I, 1428 Buenos Aires, Argentina}

\date{today}

\begin{abstract}
We analyze the effects that general environments, namely ohmic and
non-ohmic, at zero and high temperature induce over a quantum
Brownian particle. We state that the evolution of the system can be
summarized in terms of two main environmental induced physical
phenomena: decoherence and energy activation. In this article we 
show that the latter is a post-decoherence phenomenon. As the energy is
an observable, the excitation process is a direct indication of
the system-environment entanglement particularly useful at zero
temperature.  
\end{abstract}

\pacs{03.65.Bz;03.70+k;05.40+j}

\maketitle

\newcommand{\beq}{\begin{equation}}
\newcommand{\eeq}{\end{equation}}
\newcommand{\dalam}{\nabla^2-\partial_t^2}


\section{Introduction}

The open quantum system approach to the dynamic of a 
particle coupled to a reservoir provides a number 
of very interesting results. Some  are the appearence 
of noise and dissipation, decoherence 
(for a complete overview see Ref.\cite{Paz_Zurek} and references
therein), entanglement, and energy exitation \cite{nagaevEPL}. Thus, 
a necessary but not suffice condition for decoherence is the entanglement 
between the main system and the bath. As long as the composed system is governed by
a reversible time evolution, entanglement alone can not generate
decoherence. The non-unitary evolution that follows the tracing
out of the environmental degrees of freedom provides the essential
source of decoherence and dissipation. The entanglement assures
that the system measures or saves the state of the environment or
alternatively, there is information about the system state stored
in the environment. In this framework, we stress that the system-environment
interaction is more than just a matter of renormalization.

There are simple examples in which decoherence is generated by a
reservoir even at zero temperature \cite{PLA}. In general, a small
system coupled to an environment fluctuates even in the zero-T
limit. These fluctuations can take place without generating an
energy trace in the bath. The fluctuations in energy of the small
system are a peculiar fact of the entanglement with the quantum
environment \cite{buttikerPRL}.
 
In Ref.\cite{pre} we have studied the evolution of a simple
time-independent bistable system, by following the quantum
evolution of a particle initially localized at one of the minima
of a double-well potential, when coupled to an external
environment at both zero and high temperature. The zero
temperature case shows subtly different and, in some ways,
unexpected properties. Tunneling is undoubtedly quickly
suppressed, as can be seen by inspecting either the probability of
the particle to remain on the original well or the evolution of
its Wigner function. In both cases we observed typical classical
features since very early times. At zero temperature, the quantum
fluctuations of the environmental oscillators, absent in a purely
classical evolution, lead to non-zero diffusive terms. Their
effect is felt primarily through the anomalous diffusion
coefficient $f(t)$ that can have a large magnitude. We have
conjectured that these non-trivial diffusion effects induced by
the quantum environment are large enough to excite the particle
over the potential barrier. This is to be contrasted with the case
where the environment is classical forbidding any kind of
activation phenomenon. Though the late time evolution in the
presence of a quantum vacuum is in nature very different from
high-T thermal activation, we suggested that it could still be
interpreted in terms of a purely classical setting, if the
environment oscillators are described by a particular non-thermal
statistical state \cite{pre}.

The importance of general environments at zero temperature lies in
its leeway to model various physical situations. The environments
we shall study hereafter are known
as supraohmic or subohmic depending on the form of the spectral
density in the low frequency part of the spectrum. The ohmic
environment is the most studied case in the literature and produces
a dissipative force that in the limit of the frequency cutoff
$\Lambda \rightarrow 0$ is proportional to the velocity. The
supraohmic case, on the one hand, is generally used to model the
interaction between defects and phonons in metals \cite{legget} and
also to mimic the interaction between a charge and its own
electromagnetic field (see for example \cite{sonnentag}). In particular, 
the use of the supraohmic case will allow us to establish a 
close relation with the decoherence
process in quantum field theory \cite{fer}. On the other hand,  the
quantum behaviour of  ``free" electrons in mesoscopic systems is
affected by their interaction with the environment, which, for
example in such cases, consist of other electrons, phonons, photons
or scatterers. Which environment dominates the destruction of the
interference phenomena generally depends on the temperature. For
instance, the temperature dependence of the weak-localization
correction to the conductivity reveals in metals that
electron-electron interactions dominate over the phonon contribution
to decoherence at the low temperature regime.

In \cite{PLA} we analyzed the effect of quantum fluctuations
of an ohmic environment as a source of decoherence. Therein, we
presented the analytical expressions of the diffusion coefficients
at zero temperature for different physical situations and showed
that decoherence
 at zero temperature does occur contrary
to what is most commonly believed. However, the suppression of the
interferences is not as fast as it is at high temperature limit.
 In the latter case, it is
expected to happen at times of ${\cal O}(1/2M\gamma_0k_BTL_0^2)$
while we have shown that at zero temperature it is smaller than
${\cal O}(1/\gamma_0)$ \cite{PLA}.

In this article, we follow the investigation initiated in
\cite{pre}, in order to thoroughly analyze the effects that
general non-ohmic environments (in Ref.\cite{pre} only the ohmic
case was considered) at all temperatures induce over a quantum
 particle in a harmonic
oscillator potential (Quantum Brownian Motion [QBM] model). We are 
interested in analyzing the energy excitation
process for ohmic and non-ohmic environments at zero and high temperature.
  A system entangled with environmental states has a number of
properties which distinguishes it from systems for which the
ground state factorizes. In order to understand the physics
underlying the entanglement process, it is important to consider
the evolution of the energy of the main system. The energy is
always an observable, and at zero temperature, fluctuations in the
energy of the main system
 are a direct
indication of the system-environment entanglement. If the system
is isolated, it is in the lowest energy state.  Quantum
fluctuations are determined only by the diagonal elements of the
density matrix. Therefore, it is not obvious that one can make any
statement on entanglement. In general, additional information
about the non-diagonal elements is needed. If we find the main
system in an excited state, one can conclude that it is entangled 
\cite{buttikerPRL}.
Hence, in this paper we shall analyze the evolution of the mean
energy of a quantum system, coupled to an environment at zero and
high temperature, and show that the system is
energetically activated due to the coupling with the bath. In the
zero-temperature case, we shall show that there is a quantum
effect, analogous to the thermal activation process, by which
diffusion produces an increase of the energy as a function of
time; i.e. a noise-induced activation. We shall pay attention to
these effects in general environments.
The isolated harmonic oscillator (in its ground state) obeys two
important properties: minimum uncertainty and equipartition of
energy between average kinetic and potential energies. When we
study the dynamics of quantum open systems, the effect of the
environment on the system is manifested through violations of
these properties. As the energy of the subsystem is an observable,
it illustrates the distinction between separable and entangled
states. We shall show that the main system can be found in higher
energy states, no matter how weakly coupled to an environment at
zero temperature it might be.

Our main scope is to show the existence of ``noise-induced
activation" phenomenon at zero temperature, and confirm it is a
post-decoherence process. Its analogous in the high temperature
limit is the ``thermal activation" process. Even though this
phenomenon is worldwide accepted, it has not been studied for
non-ohmic environments nor has it been shown to be a
post-decoherence process. Thus, we shall analyze two different
thermal regimes and study whether  a system coupled to an
environment at arbitrary temperature, apart from suffering the
destruction of its interferences and dissipation, can benefit in
some way. We shall confirm the existence of ``thermal activation"
for non-ohmic ``decoherent" environments. In the case of zero
temperature, we are interested in answering if the
 system can increase its own energy by interacting with
 the environment
in the same way that at high temperature it is ``activated" by the
presence of it. This process has been shown to exist
 in the case of a quantum particle
localized in one minimum of a double-potential well in
Ref.\cite{pre}.

\section{The model}
\label{masterT0}

Let us consider a quantum particle (characterized by its mass $M$
and its bare frequency $\Omega$) bi-linearly coupled to an environment
composed of an infinite set of harmonic oscillators (of mass $m_n$
and frequency $\omega_m$) (for a general presentation of the model 
see, for example \cite{Grabert, Weiss,Unruh} and references therein).

The dynamics of the quantum Brownian particle  can be obtained by
tracing over the degrees of freedom of the environment and
obtaining a master equation for the reduced density matrix of the
system, $\rho_{\rm r}(t)$. We shall assume that the initial states
of the system and environment are uncorrelated, with the latter
being in thermal equilibrium at temperature $T$ (strictly zero
temperature also allowed) for $t=0$ (i.e when the interaction
between system and environment is switched on). At the initial
time, the state is a product of a given state of the system and a thermal 
state for the environment. Only
when the interaction is turned on the system is allowed to evolve.
The initial condition is not an equilibrium state of the complete
action \cite{pre}. Under these assumptions, and using that the
system-environment coupling is small, the reduced density matrix
satisfies the following master equation (we set $\hbar=1$)
\begin{eqnarray}
i\frac{\partial}{\partial t} \rho_{\rm r}(x,x',t) &=&
\Bigg[-\frac{1}{2 M^2} \bigg( \frac{\partial ^2}{\partial x^2} -
\frac{\partial ^2}{\partial x'^2} \bigg)\Bigg] \rho_{\rm
r}(x,x',t)
+ \frac{1}{2}M \Omega^2 (x^2-x'^2)  \rho_{\rm r}(x,x',t)
\nonumber \\
&+& \frac{1}{2}M {\delta \Omega}^2(t) (x^2-x'^2) \rho_{\rm
r}(x,x',t)
- i \gamma(t) (x-x') \bigg(\frac{\partial }{\partial x}
-\frac{\partial }{\partial x'} \bigg) \rho_{\rm
r}(x,x',t)\nonumber \\
&-&i M {\cal D}(t)(x-x')^2 \rho_{\rm r}(x,x',t)
-f(t)(x-x')\bigg(\frac{\partial }{\partial x} + \frac{\partial
}{\partial x'} \bigg) \rho_{\rm r}(x,x',t).
\label{master}\end{eqnarray} 
In this expression, $\delta \Omega^2(t)$ is the shift in the system frequency, which
produces the renormalized one $\tilde \Omega^2 = \Omega^2 + \delta
\Omega^2$,
 $\gamma(t)$ is the dissipation coefficient, and ${\cal D}(t)$ and
 $f(t)$ are the normal
and anomalous diffusion coefficients respectively, which produce
the decoherence effects. These coefficients have all
been defined in \cite{PLA,hpz}. The diffusion coefficients are deduced
from the noise kernel, source of stochastic forces in the
associated Langevin equation, whereas $\delta \Omega^2(t)$ and
$\gamma(t)$
 are related to the dissipation kernel. The dissipation
$\eta(t)$ and noise $\nu(t)$ kernels  are respectively defined as
$\eta (t) = \int_0^\infty d\omega I(\omega ) \sin \omega t$ and 
$\nu (t) = \int_0^\infty d\omega I(\omega ) \coth {\frac{\beta
\omega}{2}} \cos \omega t$.
$I(w)$ is the spectral density of the environment defined as $
I(\omega ) = (2/\pi) M \gamma_0 \omega(\omega/\Lambda) ^{n-1}
e^{-\omega^2/\Lambda^2}$ and $\Lambda$ is the physical
high-frequency cutoff, which represents the highest frequency
present in the environment, and $\beta=1/T$ its inverse
temperature (with Boltzmann constant set to unity, $k_B=1$). We shall 
consider the quantum system coupled to a general environment,
namely ohmic ($n = 1$), supraohmic ($n = 3$), and subohmic ($n = 1/2$).

We numerically solve Eq.(\ref{master}) considering its
coefficients up to second
order in the system-environment coupling, this fact has been taken
into account in all the simulations we shall present. We will work
in the under-damped case, which ensures the validity of the
perturbative solutions up to the times we are interested in
\cite{hpz,diana_jpp}. Hereafter, let us envisage the situation in
which $\gamma_0 \ll \hbar$, which is called the weak-interaction
situation and sets the temporal domain for perturbative solutions.
 All the results obtained below are for periods
of the evolution well within the regime for which this
approximation is valid. It is worth noting that
Eq.(\ref{master}) is valid at any temperature, and is local in
time, despite the fact that no Markovian approximation was
explicitly taken. In the following, we shall show how the
general master equation simplifies in different regimes, making it
more tractable for both analytical and numerical techniques.

In particular, in the high temperature limit, i.e. $\hbar \omega
\ll k_B T$, of an ohmic environment, the coefficients 
 of the master equation (Eq.
(\ref{master})) are constant, and therefore, the expression is
further simplified \cite{pre,hpz}. In such a case, the diffusion
coefficient can be approximated by ${\cal D}\simeq 2 \gamma_0
k_{\rm B} T M$, where $\gamma_0$ is the dissipation coefficient
\cite{hpz}. In this limit, while $\gamma_0$ is a constant and
${\cal D}(t)\propto T$, the coefficient $f \propto T^{-1}$ can be
neglected. Therefore, the term proportional to ${\cal D}$ is the
relevant one in the master equation at high temperatures in order
to evaluate, for example, the decoherence time.

On the opposite thermal regime, i.e. strictly zero temperature,
the master equation is much more complicated because the coefficients
are not constant and depend explicitly upon the time. In
\cite{PLA}, we have computed those coefficients for a quantum
Brownian particle coupled to an ohmic environment at strictly zero
temperature.

\section{Decoherence}
The decoherence process in the limit of high
temperature has been extensively studied
in the literature \cite{Paz_Zurek,hpz,phz}. However,
no much has been said about general environments,
namely supraohmic and subohmic, particularly if the environment
is at stricly zero temperature.
Therefore, as we have done in \cite{PLA} for 
an ohmic environment, in this Section
we shall analyze the decoherence process for non-ohmic 
environments 
since it is necessary for the
understanding of the excitation induced phenomena.
 We shall study the dynamic
evolution of an initial superposition of two 
delocalized (separated a distance 2 $L_0$ in position) 
states when the system is coupled to an non-ohmic
environment at zero temperature. We consider two wave packets
symmetrically located in phase space as in \cite{PLA,phz} 
and evaluate the coeficient
$A_{\rm int}$, which
results crucial to estimate the decoherence time $t_{\cal D}$
\begin{equation}
\dot{A}_{\rm int}\approx 4 L_0^2 {\cal D}(t) - 2 f(t)
,\label{aint}\end{equation}
where ${\cal D}(t)$ and $f(t)$ are the corresponding normal and
anomalous diffusion coefficients, respectively for each environment
considered.

The dynamics of the evolution of a system coupled to a supraohmic 
environment is 
quite peculiar for that everything that
happens in the system is during the initial jolt timescale \cite{hpz}. 
It is important to deepen the study of
this kind of environment since it can be very useful, for example,
as a toy model to understand the physics of interacting quantum
fields. In Ref.\cite{phases} we have made
analytical estimations of the decoherence time for this kind of
environment in a simplified spin-boson model. Notably, timescales 
obtained
therein coincide with the ones obtained numerically in the present
Letter. 

The analytical estimation of the fringe visibility factor
$A_{\rm int}$ can be
obtained for different physical limits. For times such 
that $\Omega t \ll 1$ 
(short times) it can be proved that 
$D(t)_{n=3} \sim (2M\gamma_0)/(\pi\Lambda^2)\Omega^4 t$ and 
$f(t)_{n=3} \sim - (2\gamma_0)/(\pi)\Omega t$. 
Using these expressions and  Eq.(\ref{aint}) we obtain 
$\dot {\rm A}_{\rm int} \sim 4 \gamma_0\Omega t 
(1 + 2 L_0^2 M\Omega^3/\Lambda^2)$. Thus, it is 
possible to get 
\begin{equation}
{\rm A}_{\rm int} \sim 2\frac{\gamma_0}{\Omega}\left(1
 + 2L_0^2M \frac{\Omega^3}{\Lambda^2}\right)\Omega^2t^2,
\end{equation}
which is smaller than unity (particularly
in the case $\Lambda \gg \Omega$). This means that the
decoherence factor $\Gamma(t)=\exp(-A_{\rm int})$
 \cite{matterwave} is not a decaying 
function with time and decoherence
shall not be effective in this case.

On the other side, if we ask $\Lambda t \gg 1$ and $\Omega t \geq 1$,
it can be checked that Eq.(\ref{aint}) reads 
$\dot {\rm A}_{\rm int} \rightarrow 0$, 
since both diffusion coefficients 
approach to zero so rapidly (it can be shown 
that $D(t)_{n=3} \sim 2M\gamma_0\Lambda 
\cos (\Lambda t)/\Lambda t$, and $f(t)_{n=3} 
\sim \gamma_0$). Thus,  the
fringe visibility $\rm A_{\rm int}$ is a constant 
and can be approximated by 
$\rm A_{\rm int} \approx 2 M L_0^2 \gamma_0$
 (the value for larger times in order to assure continuity
of the coefficient). 
It is easily deduced that decoherence
never occurs for this case, except for unrealistic values of the
coupling constant (values outside the perturbative treatment. For more 
general solutions at larger times see \cite{hu07}). 
The decoherence factor shall be a constant value as $\Gamma \sim
e^{-2ML_0^2\gamma_0}$ for all times. Since we are considering the
underdamped case (small $\gamma_0$), the exponent shall not be of
order one and $\Gamma(t)$ shall not be much smaller than unity. Decoherence 
shall be only effective as long as $ML_0^2 \geq 1/\gamma_0$, i.e. 
large macroscopic trajectories. This
result is to be contrasted to the one obtained in the high
temperature limit, where decoherence occurs in a time estimated as $
t_D ^{n=3,HT}\sim (\Lambda M k_B T L_0^2\gamma_0)^{-1/2}$ very
similar to what occurs in the ohmic environment at the same
temperature \cite{phases}.  

\begin{figure}[!h]
\includegraphics[width=16cm]{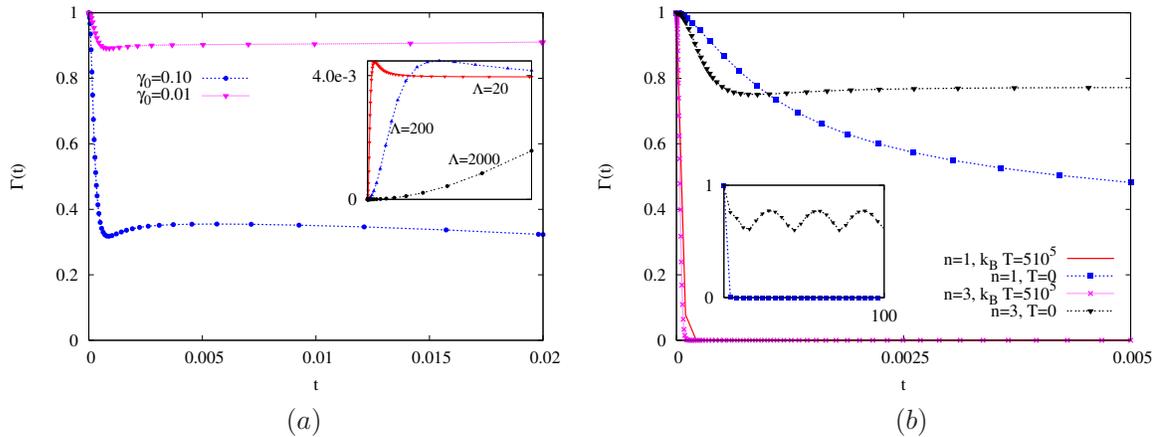}
\caption{(a) Decaying factor $\Gamma(t)$ for a supraohmic environment
($n=3$) at zero temperature. The parameters are measured in units
of the bare frequency $\Omega$. We consider the cases
$\Lambda=2000$ ($\Lambda=200$ when noted), $\Omega=0.1$, $L_0=2$
for different values of $\gamma_0$ of the environment. In the
inset, we have plotted the time evolution of the linear entropy
$Sl(t)$ for different values of the cutoff $\Lambda$ and
$\gamma_0=0.5$. There is no total decoherence for the underdamped
supraohmic
environment. (b) A comparison between different decoherence rates for
ohmic and supraohmic environments at zero and high temperature is
shown in the main plot. Decoherence is as fast in the ohmic case
as in the supraohmic for the high temperature limit, while for
those times there is no decoherence at zero-T. In the inset, the
zero-T behaviour of both environments for longer times. There is
not decoherence for the supraohmic environment, while the ohmic
takes longer but in the end, it succeeds in destroying the
interferences of the main system. Parameters used: $\Lambda=2000$,
$\gamma_0=0.1$, $L_0=1$ and are measured in units of $\Omega$.}
\label{fig1c}
\end{figure}

In Fig.\ref{fig1c}(a), we show the behaviour of the decoherence
factor $\Gamma(t)$ for two different values of the coupling
constant in the case the supraohmic environment is at zero
temperature. As expected,  the stronger the coupling  with the
environment the
 sooner $\Gamma(t)$ decreases. However, as can be seen in
Fig.\ref{fig1c}(a) it never reaches zero. This is so
because the stronger the environment,
 the bigger the initial jolt and the more efficient the
 suppression of the interferences results (the diffusion
coefficient is proportional to $\gamma_0$). For example, in the
case of
 $\gamma_0=0.01$
  in Fig.\ref{fig1c}(a), we see that $\Gamma(t)
 \approx 0.9$ after the initial transient and remains steady for
 all times. Then, interferences are not completely
 destroyed, just  slightly attenuated.
Decoherence is definitely not effective at zero-T
 in the supraohmic case for the  set of parameters chosen.
 Yet more, it is 
possible to see that the linear entropy has
 an initial jolt for all values of the
frequency cutoff and reaches an asymptotic limit (sooner for
smaller values of the frequency cutoff $\Lambda$).  However,
 we can also observe that the
linear entropy does not significantly  increase (reaches an upper
limit but far from the maximum possible value for a mixed state)
unlike in the ohmic case.

It is important to stress that in the examples shown in the plots, we have 
used parameters such that decoherence is essencially absent. We have 
ommited to show examples of macroscopic trajectories (very large $L_0$) 
since they are not of much interest from the microscopic point 
of view. Quantum interference between macroscopically distinguishable 
trajectories are easily destroyed even for the supraohmic environment 
at zero temperature.

The non-dissipative character of the supraohmic environment is a
consequence of the weakness of the spectral density in the
infrared sector, and the dependence with $\Lambda$ is due to the
fact that it is more sensible to the ultraviolet cutoff of the
frequency spectrum. The supraohmic QBM
 model can be viewed as a
toy model for a quantum field theory (QFT) scenario. In Ref.\cite{fer}
has been shown the conditions under which there is decoherence at
$T=0$ for a non-linear interacting field.  The supraohmic case is
weakly diffusive due to the fact $f(t)$ goes to zero after the
initial transient. In this case,  decoherence, depends strongly
on the coupling constant between
system and bath in order to generate remarkable effects.
In QFT, diffusive effects come from the particle
creation in the environment due to the interaction with the system. When
there is a frequency threshold in the environment, only those modes in the
system with frequency near the threshold are able to create particles
and decohere. This is why the diffusion coefficient is different from
zero only for particular values of the parameters.  This result is similar
to what we have shown so far for the supraohmic environment. In the
QBM model we are studying here, the relation between
$\Omega$, $\Lambda$, and
$\gamma_0$ is crucial in order to get diffusive effects.
Particularly at $T=0$, when
$\Omega \ll \Lambda$ the system is unable to excite the environment
in order it ``create'' particles \cite{fer}.

In conclusion, whereas supraohmic high-T environments are very
efficient inducing decoherence on the main system under certain
conditions \cite{hpz,phases}, at zero temperature there is a
strong condition over the dissipative constant $\gamma_0$. As
expected, bigger $\gamma_0$ implies a stronger coupling to the
external environment and therefore, shorter decoherence times.

Finally, in Fig.\ref{fig1c}(b), we present a comparison between the
decoherence rates for ohmic and supraohmic environments at zero
and high temperature. Decoherence is as fast in the ohmic case as
in the supraohmic for
the high temperature limit. At zero-T there is not decoherence for
the supraohmic environment, while the ohmic environment takes
longer times but finally succeeds in destroying the quantum
interferences, as can be seen in the inset of the figure where the
time scale is longer than the one of the main plot.

As for the subohmic environment, we also perform an analytical
estimation of the decoherence
timescale by computing the fringe visibility factor. In the case
that $\Lambda t \gg 1$ and $\Omega \geq 1$, Eq.(\ref{aint}) reads
\begin{equation}
\dot {\rm A}_{\rm int} \sim \gamma_0 \Lambda \left(2 M L_0^2+
\frac{\Gamma_{\rm Euler}}{\Omega}+ \frac{\log(2 \Lambda t)}{\Omega}\right).
\end{equation}
In this case, by integrating in time the above equation, we can obtain the fringe
visibility factor $\rm A_{\rm int} \sim  \gamma_0 \Lambda t/\Omega 
\log(2 \Lambda t)$. It is easy to deduce the decoherence timescale as
$ t_D \leq \Omega/(\gamma_0\Lambda)$.
On the other hand, if we ask $\Omega t \ll 1$, we would 
obtain a similar decoherence
time since the diffusion coefficients presented for the subohmic
environment depend slightly on the ratio $\Omega /\Lambda$. 
It is important to note that in order 
to be able to neglect the initial transient, one needs to have  
product $\Omega /\gamma_0 > 1$.

\begin{figure}[!ht]
\includegraphics[width=16cm]{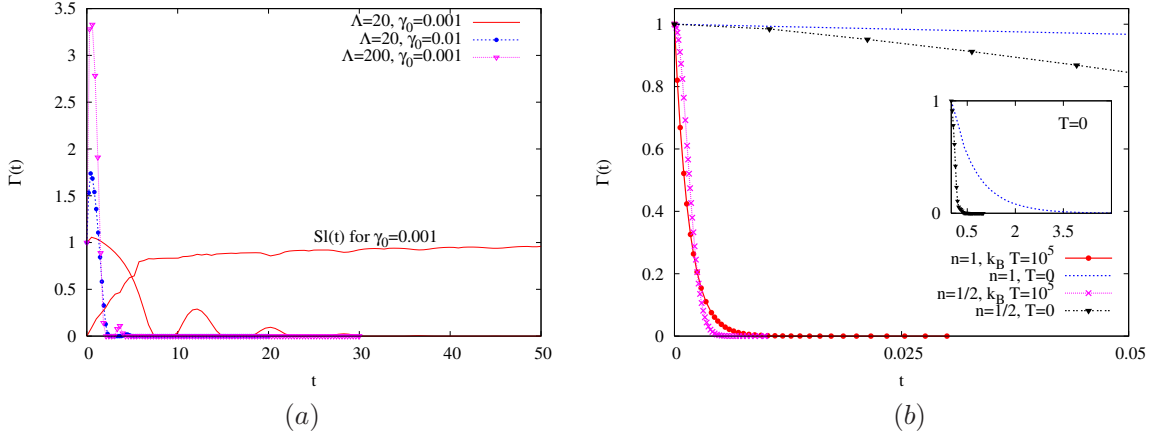}
\caption{$\Gamma(t)$ for the case of a system coupled to a
subohmic environment ($n=1/2$) at zero temperature. Parameters are
in units of the bare frequency $\Omega$. We consider the case
$\Omega=0.1$, $L_0=2$ for different values of $\gamma_0$ and the
frequency cutoff $\Lambda$. It is also shown the linear entropy
$Sl(t)$ for the indicated case. (b) A comparison between different
decoherence rates for
ohmic and subohmic environments at zero and high temperature is
shown in the main plot. Decoherence is as fast in the ohmic case
as in the subohmic for the high temperature limit, while for
those times there is yet no decoherence at zero-T. In the inset, the
zero-T behaviour of both environments for longer times.  The
decoherence timescale is shorter for the subohmic case at strictly zero
temperature. Parameters used: $\Lambda=200$,
$\gamma_0=0.01$, $L_0=0.1$ and are measured in units of
$\Omega$.} \label{fig2c}
\end{figure}

We can check our estimations with the help of
Fig.\ref{fig2c}(a), where we present the evolution of
$\Gamma(t)$ for a system coupled to a subohmic environment at zero
temperature. There is a peculiar feature in this case. The
exponential $\Gamma(t)$ initially grows but immediately after,
decreases and reaches zero \cite{hpz}.
We can easily note that the dependence with the coupling constant
and the frequency cutoff is similar to the ohmic case (the bigger
$\Lambda$ and $\gamma_0$, the sooner $\Gamma(t) \rightarrow 0$)
 in the high temperature limit. In
Fig.\ref{fig2c}(a) we present the time dependence of the
linear entropy $S_l(t)$ for just one curve (for the sake of
clarity), where it is easy to observe that the saturation of this
quantity is reached  (the asymptotic limit closer to one when the
corresponding solid curve is closer to zero and total decoherence
is effective).

Finally, a very distinctive difference between this type of
environment and the ohmic one  at zero temperature is that the
former is much more effective in producing decoherence and does so
in a shorter decoherence timescale. In Fig.\ref{fig2c}(b), we show a comparison
between the ohmic  and subohmic environments at zero temperature and in
the high temperature limit. Therein, it is easy to see, that the subohmic environment
is very efficient in inducing decoherence on the quantum test particle not
only at high temperature but at strictly zero temperature as well.

\section{Noise-induced energy excitation in the high temperature limit}
\label{energy}

We shall start by studying the thermal activation process in a
general environment, either ohmic or non-ohmic. For each case,
 we shall provide analytical arguments and
numerical evidence of the existence of this phenomenon  and
observe that, those systems which are mostly activated, are those
whose interferences have already been suppressed by decoherence in
a previous timescale.

For all cases, we shall evaluate
\begin{equation}
\langle E(t)\rangle = \frac{1}{2M}\langle p^2\rangle (t) +
\frac{M{\tilde \Omega}^2(t)}{2}\langle x^2\rangle (t) ,\nonumber
\end{equation} where $\langle x^2\rangle = {\rm Tr}(\rho_{\rm r}(t) x^2)$
and $\langle p^2\rangle = {\rm Tr}(\rho_{\rm r}(t) p^2)$ are calculated
using the solution of the master equation for a superposition 
of two Gaussian wave packets, initially
located at $x=\pm L_0$. In the following, we
 shall present the
mean energy of the system for different environments at all
temperatures so as to show the ``noise-induced activation"  in a
quantum brownian particle.

\subsubsection{Ohmic Environment}

It is widely accepted that ``thermal" activation is a
post-decoherence process for a system coupled to an ohmic
environment in the high temperature limit. This is so because,
after the suppression of the system's interferences due to the
presence of the environment, the system and the environment are
still in interaction. Therefore, there is still energy exchange
between them and consequently, the system increases its energy proportionally
to the temperature of the bath for short times. The thermal
activation rate for a classical system can be obtained by working
with the classical analogue of the master equation for the Wigner
function, the Fokker-Planck equation:
\begin{eqnarray}
\dot{W}=\{H_{\rm sys},W\}_{\rm PB} +2 \gamma_0 \partial_p(pW) +
{\cal D}
\partial^2_{pp}W\,. \label{fp}
\end{eqnarray} 
The classical evolution for the average of any
physical observable $A(x,p)$ in this regime is then given by:
\begin{equation}
\partial_t \langle A \rangle = -\langle\{H_{\rm sys},A\}_{\rm PB}
\rangle +
 {\cal D} \langle \partial^2_p A \rangle - 2 \gamma_0 \langle p
\partial_p A\rangle.
\end{equation}

\begin{figure}[!ht]
\includegraphics[width=16cm]{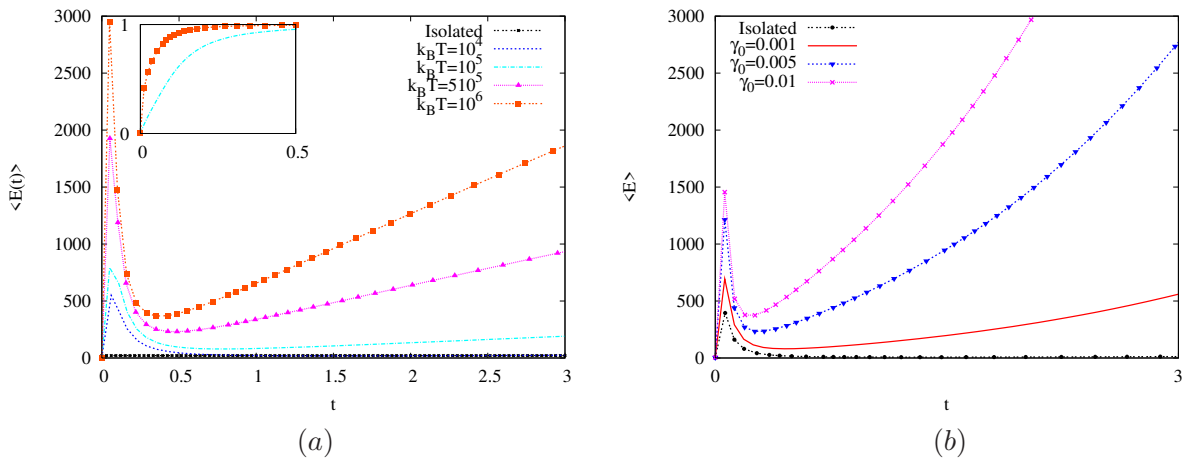}
\caption{(a) Evolution in time of the mean energy of the system
coupled to an ohmic environment ($n=1$) in the high temperature
limit.  We consider the case $\gamma_0=0.001$, $\Lambda=2000$,
$\Omega=0.1$, $L_0=2$ for different temperatures of the environment.
In the inset, we have plotted the time evolution of the linear
entropy for two different environments: $k_B T=10^5$ and $k_B
T=10^6$ (same colors than in the main picture). Thermal activation
is a post-decoherence phenomena. (b) Evolution in time of the mean
energy of the system coupled to an ohmic environment ($n=1$) in the
high temperature limit. We consider the case $k_BT=10^5$,
$\Lambda=2000$, $\Omega=0.1$, $L_0=2$ for different values of the
coupling constant $\gamma_0$.  Big values of $\gamma_0$ activate
the system earlier. The parameters are measured in units
of the bare frequency $\Omega$ in all cases.} \label{fig3c}
\end{figure}

If we take $A(x,p)$ to be the Hamiltonian of the main system, we
obtain $\partial_t \langle H \rangle = 2 \gamma_0 (k_BT - \langle
p^2 \rangle)$ (being ${\cal D}=2 \gamma_0 k_BT $ for this case).
This expression can be further simplified by assuming $T$ to be
much higher than the relevant energy scales in the problem,
$\langle p^2 \rangle$, during the early stages of the evolution.
As a result, the time dependence of the energy of the system is
given by:
\begin{equation}
\partial_t \langle H \rangle =
2 \gamma_0 k_B T \,\,\,\rightarrow\,\,\, E= 2 \gamma_0 k_BT t +
E_0~, \label{EHT}
\end{equation}
where $E_0$ is the initial energy of the system. We can then
estimate the thermal activation time $t_{\rm th}$
 as $t_{\rm th}= (E-E_0)/(2 \gamma_0 k_BT)$. In 
Fig.\ref{fig3c}(a) we have plotted the time evolution of the
system's mean energy for the ohmic environment and confirmed its
behavior is proportional to the temperature of the bath as indicated
in Eq.(\ref{EHT}). We can note that, initially in the cutoff
timescale, the energy develops a jolt. This is just a transient and
does not last long. The energy grows steadily for a while
(proportional to $\gamma_0 k_B T $ for short times) and does not
depend on the frequency cutoff $\Lambda$. It is important to note
that we shall always study the dynamics between times $1/\Lambda \ll
t \ll t_{\rm sat} \sim 1/\gamma_0$. As it is already known \cite{hpz}, the
decoherence timescale is $t_D\sim 1/(2 M \gamma_0 k_B T L_0^2)$
(very early for the environments of Fig.\ref{fig3c}(a)).
  In the inset
of that figure, we can see the timescale at which the linear entropy
reaches its top value (unity for a mixed state). 
By this time, interferences have already been
destroyed. Besides,
 the energy of the open system is always bigger
than the one of the closed one.  Note that
 not only the decoherence process but also the noise
activation of the system depend strongly on the external temperature
of the environment. The hotter the environment, the bigger the
``activation" (thermal in this case). In Fig.\ref{fig3c}(b) we can
check its dependence on the value of the coupling constant
$\gamma_0$ at a fixed temperature and cutoff frequency.  As
expected, the bigger the coupling constant $\gamma_0$, the bigger
the increase of energy of the system. In Fig.\ref{fig3c}(b) it is
shown that the coupling strength sets the time at which the energy
starts growing. The long time evolution of the energy in the 
high temperature case can be found in Ref. \cite{pre}.

\subsubsection{Nonohmic Environments}

The previous analysis can be also done for non-ohmic
environments. However, expressions are not that easy to deal with
since, in this case, neither $f(t)$ is negligible nor ${\cal D}$
constant. Thus, we shall restrict ourselves to a numerical
analysis based on the thorough analytical study of the decoherence
process we have made in a previous Section.

\begin{figure}[!ht]
\includegraphics[width=16.5cm]{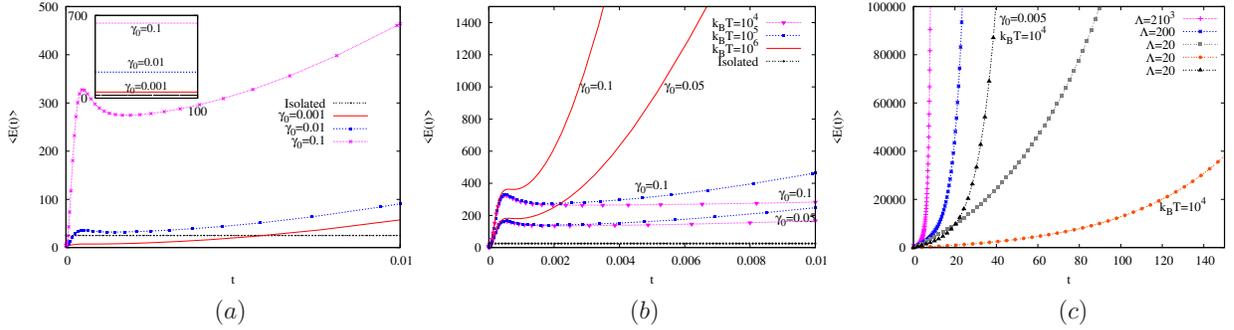}
\caption{(a) Evolution in time of the mean energy of the system
coupled to an supraohmic environment ($n=3$) in the high temperature
limit. We consider the cases $k_BT=10^5$, $\Lambda=2000$,
$\Omega=0.1$, $L_0=2$ for different values of $\gamma_0$. In the
inset, we have plotted the
 mean energy for longer times. (b) Evolution in time of the mean energy of the system
coupled to an supraohmic environment ($n=3$) in the high temperature
limit.  We consider the cases $\gamma_0=0.001$, $\Lambda=2000$,
$L_0=2$ for different temperatures of the environment and coupling
constants $\gamma_0$. Parameters are measured
in units of the bare frequency $\Omega$ in all cases.
(c) Evolution in time of the mean energy of the
system coupled to an subohmic environment ($n=1/2$) in the high
temperature limit $k_B T>> \hbar \Lambda$.  We consider the cases
$\Omega=0.1$, $L_0=2$, $\gamma_0=0.001$ (and $\gamma_0=0.005$ when
noted) and $k_B T=10^5$ ($k_B T=10^4$ when noted) for different
values of the frequency cutoff $\Lambda$. Parameters are measured
in units of the bare frequency $\Omega$ in all cases.} \label{fig4c}
\end{figure}

We begin our analysis with the supraohmic one. In Fig.\ref{fig4c}(a)
and Fig.\ref{fig4c}(b) we present the evolution in time of the mean
energy of the system for different values of the coupling constant
and the external temperature respectively. In both cases, it is easy
to see that those curves where the initial jolt of the energy is
bigger are those cases for which the mean energy will be bigger in
the final state (top inset in Fig.\ref{fig4c}(a)). In particular,
therein, we can see that the mean energy is bigger as the coupling
constant grows and in Fig.\ref{fig4c}(b), we see the same behavior as
the environment gets hotter.  It is important to note that the
strength of the environment is given by a relation between the three
parameters $\gamma_0$, $k_BT$ and $\Lambda$.

For ``strong" supraohmic environments ($2Mk_BTL_0^2\gamma_0 \gg
\Lambda$), decoherence happens in timescales of
$t_{\cal D}\sim (\Lambda M \gamma_0 k_B T L_0^2)^{-1}$
 since interferences are destroyed very
rapidly.  Surprisingly, those curves correspond to a considerably
bigger value of the final mean energy of the system.

So far, we have shown that when the environment has ``succeeded"
in the destruction of the interferences, the exchange of energy
with the system promotes its ``activation". However, for not so
strong environments ($Mk_BTL_0^2\gamma_0<\Lambda$), 
such as $\gamma_0=0.001$ in Fig.\ref{fig4c}(a), the
interferences are not completely destroyed (the decoherence factor 
$\Gamma(t)$ is not
zero), and then, the exchange of energy is not completely devoted
to exciting the system (the environment still tries to suppress
the coherences). This case differs qualitatively from the ohmic
environment. In particular, it reaches an asymptotic limit in a
very short timescale. In the inset of Fig.\ref{fig4c}(a) 
we show the evolution of the 
mean energy for longer times, when the system enters in the 
asymptotic regime for the given set of parameters.

Finally, the noise activation is very clear in the case of the
subohmic environment. In Fig.\ref{fig4c}(c) we have plotted the mean
energy of the system for different values of the coupling constant
($\gamma_0=0.001$ and $\gamma_0=0.005$), different temperatures
($k_BT=10^5$ and $k_BT=10^4$) and different cutoffs in the high
temperature limit. It is easy to see that the energy grows faster as
the coupling constant, the frequency cutoff and the temperature
grow. If we recall the behaviour of the decoherence factor $\Gamma(t)$
or the decoherence timescale $t_{\cal D}\sim (M \gamma_0
L_0^2 k_B T)^{-1}$, we can check once more, that the ``noise activation" is
a post-decoherence phenomenon since it begins after the system has already lost
its interferences.

\section{Noise-induced energy excitation at Strictly Zero Temperature}

As we have already mentioned, there is a widely spread
misconception that decoherence tends to zero as a function of the
temperature and therefore, there is no decoherence in the limit of
zero environmental temperature. If that were the case,  physics
should be different in the opposite thermal limit. Many questions
consequently arise. What do we expect to find
 if the system is coupled to an ohmic environment at zero
temperature? Is it possible to find ``activation" in the system?
The latter answer is unexpectedly positive. Yet more, we shall see
that there is a close connection between decoherence and
noise activation since the most  ``decoherent" environments are
the ones with the most visible activation phenomenon. Once more,
we shall show that the latter is a post-decoherence process taking
place after the interferences have been suppressed.

When trying to interpret the post-decoherence behavior of the open
system, several features of its dynamics should be kept in mind.
Firstly, one should emphasize that the initial condition is
clearly not the ground state of the total action \cite{pre}. As
soon as the interaction between the main system and the
environment is turned on, at $t=0$, the system will find itself in
an excited energy state. The environment will have a non-zero
amount of energy in relation to the new initial state. From a
purely classical point of view, this energy cannot be responsible
for the excitation of the particle to higher values. This argument
can be made more quantitative in the following way: the full
potential for the system plus environment is
\begin{equation}
V(x,q_n) = V_{\rm sys}(x) + V_{\rm env}(q_n) + V_{\rm int}(x,q_n),
\end{equation}
with $V_{\rm sys}(x) = -\frac{1}{2} \Omega^2 x^2$, 
$V_{\rm env}(q_n)= \sum_n \frac{1}{2} \omega^2 m_n^2 q_n^2$
and $V_{\rm int}(x,q_n)= \sum_n C_n x q_n$.
Classically, the initial condition is $ x=0$,
 and, because the environment is at T=0, $q_n=0$. So, for the
 full action, the energy terms of the initial condition are
 given by $V_{\rm sys} =0$
(the minimum of $V_{\rm sys}$), $V_{\rm env}=0$, and  $V_{\rm
int}=0$. Consequently, the total initial potential energy of the
system plus environment is  $V=0$. Note that classically, the
value of the total energy is the same as the
 energy of the isolated main particle, even when the interaction
 with the environment is ``switched on". This is a consequence
 of taking zero temperature for the environment. The quantum fluctuations
present in the initial state of the environment must play a role
in the ``activation'' \cite{pre}. One should note that these fluctuations are
not ``vacuum fluctuations" of the full system. Nonetheless, the
fact that they have such a clear effect on the evolution of the
system is quite remarkable. Whereas in the high-T case the quantum
and classical oscillators composing the bath had identical
distributions, they behave in a markedly different way as
$T\rightarrow 0$. The quantum nature of the environment, which
could be ignored at high-T, leads in this limit to important
non-negligible effects.

In terms of the master equation, the quantum fluctuations of the
 bath
oscillators generate non-zero $f(t)$ and ${\cal D}(t)$ terms,
making diffusive phenomena possible. This is particularly true in
the case of the anomalous diffusion coefficient $f(t)$. In the ohmic 
case it can be shown  that it 
depends logarithmically on the cutoff $\Lambda$ and thus can be
considerably large \cite{PLA}. Diffusion
 effects induced
by quantum fluctuations  are thus responsible for exciting the
particle. Though this process is very different from high-T
thermal activation, we conjecture that it may still be interpreted
in terms of a modified classical setting. The key ingredient is
that the classical bath
 should mimic the
properties of the quantum $T=0$ environment. Considering the
classical and quantum versions of the noise kernel $\nu(s)$, it is
possible to show that a bath of classical oscillators with a
frequency dependent temperature $T(\omega)=\hbar \omega/2$ should
reproduce the effects of the initial quantum state. In fact, for
this choice of classical environment one obtains $f(t)$ and ${\cal
D}(t)$ terms identical to those of the $T=0$ quantum case. Our
main point is that after decoherence takes place, a quantum open
system at $T=0$ should behave as a classical open system in
contact with a classical bath whose oscillators are excited in a
way that reproduces the fluctuations of the corresponding quantum
environment \cite{pre}. Hereafter, we shall show noise-induced
activation happening in different environments a strictly zero
temperature.

\subsubsection{Ohmic Environments}

\begin{figure}[!h]
\includegraphics[width=16cm]{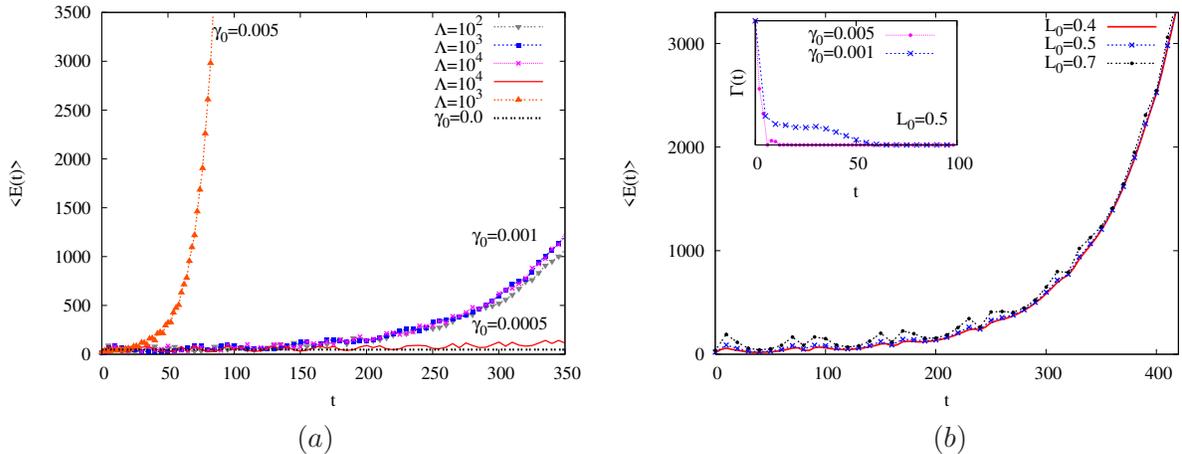}
\caption{(a) Evolution in time of the mean energy of the system
coupled to an ohmic environment ($n=1$) at zero temperature. It
is clear to see that there is a dependence upon the coupling
constant but not on the frequency cutoff. (b) Evolution in time of
the mean energy of the system coupled to an ohmic environment
($n=1$) at zero temperature. The parameters are measured in units of
the bare frequency $\Omega$. We consider the case $\gamma_0=0.001$,
$\Omega=15$, $\Lambda=1000$ for different values of  $L_0$
. In the inset, 
we show the time evolution of the decoherence factor for different values 
of $\gamma_0$. Energy excitation actually starts after decoherence 
becomes effective. The parameters are chosen based on numerical convenience. }
 \label{fig6c}
\end{figure}
In Fig.\ref{fig6c}(a) we numerically prove our statement for an ohmic
environment at zero temperature. Therein, we have plotted the time
evolution of the mean energy of the system for different values of
the frequency cutoff and the coupling constant $\gamma_0$ for an
ohmic  environment. The bigger the value of $\gamma_0$, the sooner
the energy of the system grows. Initially, the mean energy of the
system is lower than that it would be if the system was isolated (no
interaction with the environment). However, after a time 
$t \geq t_{\cal D}$, timescale for which the interferences have already been
suppressed due to decoherence, the mean energy increases
considerably because of the interaction with the environment. The
system gains energy at the expense of the environment which turns
out to be an energy source. Clearly, it is a ``post-decoherence"
process, very much like the thermal activation at high temperature.
In Fig.\ref{fig6c}(b), we have plotted the mean energy of the system
for different values of the initial distance
 between the Gaussian wave packets $L_0$ \cite{matterwave}. As
expected, the energy doesn't depend much on this latter parameter:
the bigger the initial distance $L_0$ is, the sooner the decoherence
process takes place and the ``noise activation" starts (since decoherence 
times is in general proportional to $L_0^{-2}$). But this
time difference is subtle. In the inset of Fig.\ref{fig6c}(b), we can
observe the exponential decay of $\Gamma(t)$ which indicates the
timescale at which interferences are completely destroyed. Notably,
this timescale coincides with the beginning of the ``noise-induced
activation" phenomenon.

\begin{figure}[!h]
\includegraphics[width=16cm]{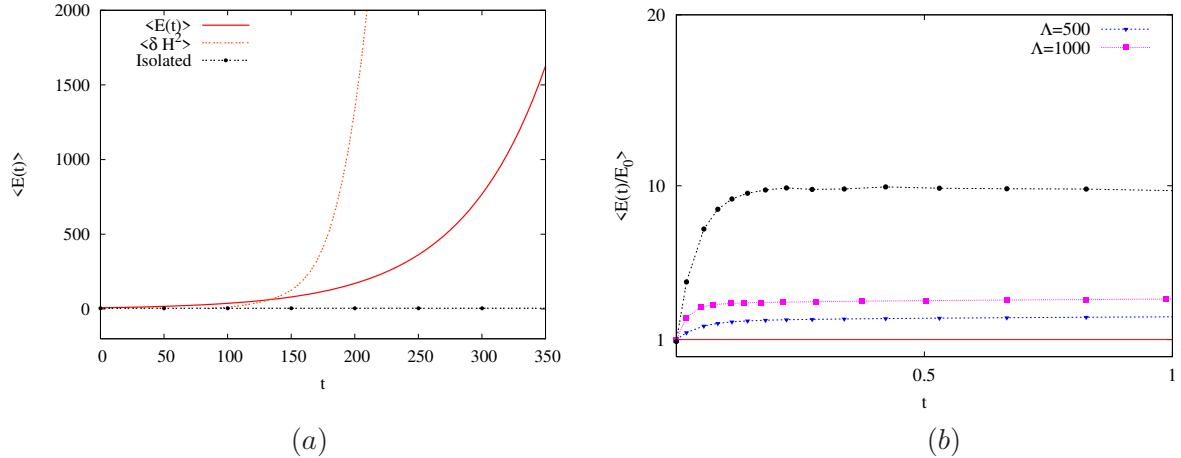}
\caption{(a) Evolution
in time for the mean energy of the system and the mean dispersion of
the energy for an ohmic environment at zero temperature. It is also shown 
the mean energy of the isolated system.
Parameters used: $\Lambda=10^3$, $\Omega=15$, $L_0=0.5$ and
$\gamma_0=0.001$. Fluctuations in energy are relevant in the
excitation process. The
parameters are measured in units of the bare frequency $\Omega$ in all cases.
(b) Long time behaviour for an ohmic environment at zero temperature 
for different values
of the parameters. In all cases the value of the mean energy of 
the Brownian particle is
bigger than the mean energy of the isolated system (solid line). The parameters used are measured
in units of $\gamma_0$ so as to include different runs in the same figure. We consider
the case $\Omega=1$ and $\gamma_0=0.1$ and different values of $L_0$: $L_0=0$ which means
only one Brownian particle (squares
and triangles) and  $L_0=0.5$ (circles). $E_0$ is the energy of the isolated system. 
Parameters were deliberate choosen due to the 
long-time difficulty in the numerical runs.}
 \label{fig7c}
\end{figure}

In Fig.\ref{fig7c}(a) we have plotted the mean energy of the
open and closed system and the mean dispersion of the energy of the
open system for zero temperature. Obviously, the mean energy of the
closed system remains steady. In the open case, we can see that,
initially, it is lower than the isolated case, but  immediately
starts growing. The dispersion of the energy shows that the
fluctuations are extremely important and are responsible for the
increase of the mean energy since it is
 an uniform growing function of time.

Finally, we would like to show the correct long time behaviour 
of the mean energy. Consequently,
 we numerically solved the master equation for times of the order of 
the saturation time $t_{\rm sat} \sim 1/\gamma_0$. 
As we are working in the underdamped limit, achieving these long timescales is a
numerically expensive task. Nonetheless, we present the correct 
asyntotic behaviour by an appropiate selection of
the physical parameters. In Fig.\ref{fig7c}(b) we show the correct long time behaviour
for the ohmic environment at zero temperature for different values
of the parameters. In all cases the value of the mean energy of the Brownian particle is
bigger than the mean energy of the isolated system (solid line).

\subsubsection{Nonohmic Environments}
\begin{figure}[!ht]
\includegraphics[width=16.5cm]{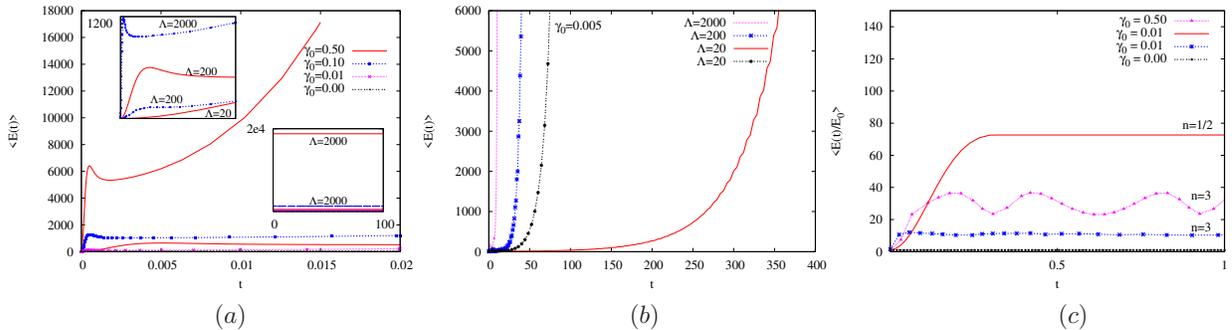}
\caption{(a)Evolution in time of the mean energy of the system
coupled to an supraohmic environment ($n=3$) at zero temperature.
The parameters are measured in units of the bare frequency $\Omega$.
We consider the cases $\Lambda=2000$, $\Omega=0.1$, $L_0=2$ for
different values of the cutoff frequency and the coupling constant
$\gamma_0$. In the top inset, we have plotted the mean energy in a
shorter scale to see the dependence on the frequency cutoff. In the
bottom inset, we have plot the mean energy for longer times for some
values of the main plot. (b) Evolution in time of the mean energy of
the system coupled to an subohmic environment ($n=1/2$) at zero
temperature. The parameters are measured in units of the bare
frequency $\Omega$. We consider the cases $\Omega=0.1$, $L_0=2$,
$\gamma_0=0.001$ for different values of the frequency cutoff
$\Lambda$. We also plotted the case $\gamma_0=0.005$ and $\Lambda
=20$ in order to compare.
(c) Long time behaviour for non-ohmic environments at zero temperature. We consider the case
of supraohmic $n=2$ and subohmic $n=1/2$ environments. The parameters used are measured
in units of $\gamma_0$ so as to include different runs in the same figure. We consider
the case $\Omega=1$ and $L_0=2$ for different values of $\gamma_0$ so as to distinguish
between ``strong'' and ``weak'' supraohmic environments. $E_0$ is the energy of the isolated system. 
Parameters were deliberate choosen due to the 
long-time difficulty in the numerical runs.} \label{fig9c}
\end{figure}

We shall extend the above analysis to the supraohmic environments.
For example, in Fig.\ref{fig9c}(a), we have plotted the time
evolution of the mean energy of the system at zero temperature.
Therein, we can see that the energy has the initial jolt but at long
time it remains constant (see bottom inset). It reaches a value and
stays with that energy for ever since much of what happens in the
supraohmic environment is just a consequence of that initial jolt,
not a dynamic response. By the way, this type of environment at 
strictly zero temperature, does
not induce effective decoherence on the system. No decoherence
implies no energy activation. 
In the Fig.\ref{fig9c}(a) we have included a big value
of $\gamma_0$ ($\gamma_0=0.5$) which produces decoherence and
therefore energy activation. However, this value is not appropriate
because it is not well covered by the underdamped approximation
(i.e. weakly coupling with the environment). All the other values of
 $\gamma_0$ included in the Figure, correspond to environments
not strong enough to neither  destroy the interferences nor
``activate" the system.

As for the subohmic environment at zero
temperature, we can observe the same dynamics as in the high
temperature limit,
 but on a longer timescale as it is shown in
 Fig.\ref{fig9c}(b).  In this case, it is also possible to check
that the activation timescale $t_{\rm act}$ is subsequent to the
decoherence timescale $t_{\rm D} \sim \Omega/(\Lambda \gamma_0)$ 
with the help of
Fig.\ref{fig9c}(b).

Finally, in Fig.\ref{fig9c}(c) we show the long time evolution of the mean energy 
for non-ohmic environments for different values of the parameters. In all
cases, we can see that an asyntotic 
behaviour is reached for the very long timescales.

\section{Final Remarks}

We have studied general environments, namely ohmic and non-ohmic
and showed that the evolution of a QBM particle can be summarized
in terms of two main environmental induced physical phenomena:
decoherence and energy activation. 

We have presented the different physical features of the
non-unitary dynamics in the case of a quantum system coupled to a
general environment at zero temperature. In this thermal regime,
the quantum fluctuations of the environmental oscillators, absent
in a purely classical evolution, lead to non-zero diffusive terms.
Their effect is felt primarily through the anomalous diffusion
coefficient $f(t)$ that can have a large magnitude in the ohmic
and subohmic environments. The supraohmic case is weakly diffusive
due to the fact $f(t)$ goes to zero after the initial transient.
In this case, both decoherence and energy excitation, depend
strongly on the coupling constant between system and bath in order
to generate remarkable effects. Yet more, we have confirmed
previous results on decoherence in quantum field theory by the use
of this model.

We have shown that these non-trivial diffusion effects induced by
the quantum environment are large enough to excite the particle to
higher energy levels. Particularly it is a post-decoherence
process which means that no total decoherence implies no
energy activation, as was clearly shown in the case of a
supraohmic environment at zero temperature. This is to be
contrasted with the case where the environment is classical
forbidding any kind of activation phenomena. Though the late time
evolution in the presence of a quantum vacuum is in nature very
different from high-T thermal activation, we suggest that it could
still be interpreted in terms of a purely classical setting, if
the environment oscillators are described by a particular
non-thermal statistical state. We will pursue this line of enquire
in depth in a forthcoming publication.

A system entangled with environment states has a number of
properties which distinguishes it from systems for which the ground
state factorizes. In order to understand the physics underlying the
entanglement process, it is important to consider the evolution of
the energy of the main system. The energy is always an observable,
and at zero temperature, fluctuations in the energy of the main
system are a direct indication of the system-environment
interaction. We have shown that the process we have called
noise-induced energy activation is a post-decoherence process,
therefore, the evolution of the linear entropy shows that the state
of the system in all of these cases is a mixed state. Thus, energy
here can be used as a meassure of the degree of entanglement for
mixed states, under a non-unitary evolution. What's more, simple
systems with well known isolated quantum mechanical properties
become ``entanglement meters" as nicely explained by Jordan and
B\"uttiker in \cite{buttikerPRL}.

\acknowledgments

This work is supported by CONICET, UBA, and ANPCyT, Argentina.

\end{document}